%% file: ECIR18_NeuralPassageModel.tex
\documentclass[runningheads,a4paper]{llncs}

\usepackage{lipsum} 
\usepackage{balance}
\usepackage{paralist}
\usepackage{filecontents}
\usepackage{etoolbox}
\usepackage{mathrsfs}
\usepackage{mathtools}
\usepackage{bm}
\usepackage{setspace}
\usepackage{amssymb}
\usepackage{graphicx}
\usepackage{wrapfig}

\apptocmd{\thebibliography}{\scriptsize}{}{}

\usepackage{url}
\urldef{\mailsa}\path|{aiqy, brenocon, croft}@cs.umass.edu|
\newcommand{\keywords}[1]{\par\addvspace\baselineskip
	\noindent\keywordname\enspace\ignorespaces#1}

\begin{document}
	
	\abovedisplayskip=0pt
	\abovedisplayshortskip=0pt
	\belowdisplayskip=0.5pt
	\belowdisplayshortskip=0pt
	\makeatletter
	\let\origsection\section
	\renewcommand\section{\@ifstar{\starsection}{\nostarsection}}
	
	\newcommand\nostarsection[1]
	{\sectionprelude\origsection{#1}\sectionpostlude}
	
	\newcommand\starsection[1]
	{\sectionprelude\origsection*{#1}\sectionpostlude}
	
	\newcommand\sectionprelude{%
		\vspace{-1.05em}
	}
	
	\newcommand\sectionpostlude{%
		\vspace{-0.7em}
	}
	\makeatother

	\mainmatter  % start of an individual contribution
	
	\title{A Neural Passage Model for Ad-hoc Document Retrieval}
	
	\titlerunning{A Neural Passage Model for Ad-hoc Document Retrieval}
	
	
	\author{Qingyao Ai \and Brendan O'Connor \and  W. Bruce Croft}
	\authorrunning{Qingyao Ai, Brendan O'Connor and W. Bruce Croft}
	
	\institute{College of Information and Computer Sciences, University of Massachusetts Amherst, 
		Amherst, MA, USA, 01003-9264 \\
		\mailsa}
	
	
	\maketitle

	\begin{abstract}
		Traditional statistical retrieval models often treat each document as a whole. In many cases, however, a document is relevant to a query only because a small part of it contain the targeted information.  
		In this work, we propose a neural passage model (NPM) that uses passage-level information to improve the performance of ad-hoc retrieval.
		Instead of using a single window to extract passages, our model automatically learns to weight passages with different granularities in the training process.
		We show that the passage-based document ranking paradigm from previous studies can be directly derived from our neural framework.
		%Our neural passage model can automatically learn to weight passages with different granularities.
		%Also, we show that several previously proposed passage-based statistical retrieval models can be directly derived from the same framework of our model with predefined structures and parameters.
		Also, our experiments on a TREC collection showed that the NPM can significantly outperform the existing passage-based retrieval models.
	\end{abstract}

	\input{introduction}

	\input{related}
	\input{model}

	\input{methodology}

	\input{experiment_results}

	\input{conclusion}

	\section{Acknowledgments}
	This work was supported in part by the Center for Intelligent Information Retrieval and in part by NSF IIS-1160894. Any opinions, findings and conclusions or recommendations expressed in this material are those of the authors and do not necessarily reflect those of the sponsor.
	
	\bibliographystyle{splncs03}
	\bibliography{sigproc}  % sigproc.bib is the name of the Bibliography in this case
\end{document}

%% file: introduction.tex
%!TEX root=ECIR18_NeuralPassageModel.tex
\section{Introduction}
\label{sec:introduction}

%ad-hoc retrieval
Ad-hoc retrieval refers to a key problem in Information Retrieval (IR) where documents are ranked according to their assumed relevance to the information need of a specific query formulated by users~\cite{baeza1999modern}. 
%Despite this simple definition, ad-hoc retrieval is a key problem in Information Retrieval (IR) and a core function of modern search engines.
%In the past decades, statistical models have dominated the research on ad-hoc retrieval and well-known examples like BM25~\cite{robertson1994some}, language modeling approaches~\cite{ponte1998language} and their variations have considerably influenced the design of modern search engines~\cite{robertson1994some,ponte1998language}.
%In the past decades, statistical models have dominated the research on ad-hoc retrieval and have considerably influenced the design of modern search engines~\cite{robertson1994some,ponte1998language}.
In the past decades, statistical models have dominated the research on ad-hoc retrieval.
They assume that documents are samples of n-grams, and relevance between a document and a query can be inferred from their statistical relationships.
%For example, the language modeling approach estimates a language model for each document with Maximum Likelihood Estimation (MLE) and ranks documents according the likelihood that the query is generated from the same language model with the document.
To ensure the reliability of statistical estimation, most models treat each document as single piece of text. 
%We refer this type of models as document-based retrieval models. 

%why passage
There are, however, multiple reasons that motivate us not to treat a document as a whole.
First, there are many cases where the document is relevant to a query only because a small part of it contains the information pertaining to the user's need. 
The ranking score between the query and these documents would be relatively low if we construct the model with statistics based on the whole documents.
Second, because reading takes time, sometimes it is more desirable to retrieve a small paragraph that answers the query rather than a relevant document with thousands of words.
For instance, we do not need a linux textbook to answer a query ``linux copy command".

%First, reading takes time. 
%It is more desirable to retrieve a small paragraph that answers the query's need than a relevant document with thousands of words with respect to user satisfaction.
%For instance, we do not need a textbook about the linux operating system to answer a query ``linux copy command".
%Second, there is a gap between document's relevance and statistic significance.
%Second, document-based retrieval models cannot capture relevance information in a local area.
%Traditional statistic models try to capture the statistic correlation between queries and the whole document. 
%In practice, there are many cases where the document is relevant only because a small part of it contains the information pertaining to the query. 
%The relevance score between the query and these documents would be relatively low if we construct the model with statistics based on the whole documents.
 
%what current model
Given these observations, IR researchers tried to extract and incorporate relevance information from different granularities for ad-hoc document retrieval. 
One simple but effective method is to cut long documents into pieces and construct retrieval models based on small passages. %, which we refer as passage-based retrieval models.
Those passage-based retrieval models are able to identify the subtopics of a document and therefore capture the relevance information with finer granularities.
Also, they can extract relevant passages from documents and provide important support for query-based summarization and answer sentence generation.
%There are at least two advantages of passage-based retrieval models.
%First, passage-based retrieval models are able to identify the subtopics of a document and therefore capture the relevance information with finer granularities.
%Second, passage-based retrieval models can extract relevant passages from the documents and provide important support for query-based document summarization and answer sentence generation.
 
%what is the problem
Nonetheless, the development of passage-based retrieval models is limited because of two reasons. 
First, as far as we know, there is no universal definition of passages in IR. %the definition for passages is not conclusive. 
Most previous studies extracted passages from documents with a fixed-length window. 
This method, however, is not optimal as the best passage size varies according to both corpus properties and query characteristics. %, which are usually unknown beforehand. % and tune it on the to find %, but we do not know the size of the best passage beforehand. 
Second, aggregating information from passages with different granularities is difficult. 
The importance of passages depends on multiple factors including the structure of documents and the clarity of queries. 
For example, Bendersky and Kurland~\cite{bendersky2008utilizing} noticed that passage-level information is not as useful on highly homogeneous documents as it is on heterogeneous documents. 
A simple weighting strategy without considering these factors is likely to fail in practice.

%our model

In this paper, we focus on addressing these challenges with a unified neural network framework. 
Specifically, we develop a convolution neural network that extracts and aggregates relevance information from passages with different sizes.
In contrast to previous passage-based retrieval models, our neural passage model takes passages with multiple sizes simultaneously and learns to weight them with a fusion network based on both document and query features.
%Compared to previous passage-based retrieval models, our neural passage model does not require a single window to extract passages. 
%Instead, it takes passages with multiple sizes simultaneously and learns to weight them with a fusion network based on both document and query features.
%In fact, our neural passage model is highly expressive as we show that the state-of-the-art passage-based retrieval models can be viewed as special cases for our model with predefined network parameters. 
Also, our neural passage model is highly expressive as the state-of-the-art passage-based retrieval models can be incorporated into our model as special cases.
We conducted empirical experiments on TREC collections to show the effectiveness of the neural passage model and visualized the network weights to analyze the effect of passages with different granularities.

%we try to address two challenges that limit the performance of passage-based retrieval models.
%The first one is how to select passage size for passage extraction.
%Most of previous studies define passage globally with arbitrary hyper-parameters and tune those parameters manually to obtain good retrieval performance.
%We believe that this method is suboptimal since we do not know the size of best passage beforehand.
%Therefore, we design a convolution neural network and can automatically extract and weight passages with different size.
%The second challenge of passage-based retrieval models is how to effectively aggregate information from text with different granularities.

%However, the definition of paragraph is vague in practice and most of previous studies rely on model hyper-parameters to specify the paragraph size in global. 
%Our argument is that this hand-tuning process is both suboptimal and unnecessary. 
%Instead, we propose a neural network model that automatically  weights paragraph with different sizes in a learning framework.
%We show that previously proposed passage-based ad-hoc retrieval models can in fact be derived from our neural framework with special network structure and predefined model parameters.   
%In addition, the flexibility of our proposed neural network framework enable us to easily incorporate word embeddings and query quality information into paragraph models and yield a more effective retrieval model that significantly outperforms the state-of-art paragraph retrieval models. 

%% file: related.tex
%!TEX root=ECIR18_NeuralPassageModel.tex
\section{Related Work}\label{sec:related}

%There are two lines of research in passage-based retrieval models that are directly related to our work.
%Previous studies mainly focus on addressing two challenges of utilizing passages for ad-hoc retrieval -- how to extract passages and how to construct passage model.
  
%pargraph size
%\subsection{Passage Extraction}
\textbf{\textit{Passage Extraction}}.
%There is no universal definition of passages in information retrieval. 
Previous studies have explored three types of passage definitions: structure-based, topic-based and window-based passages. 
Structure-based passage extraction identifies passage boundaries with author-provided marking such as empty line, indentation etc.~\cite{salton1993approaches}. 
%It provides natural segmentations for documents but fails when text data does not contain explicit boundary symbols. 
Topic-based passage extraction, such as TextTiling~\cite{hearst1993texttiling}, divides documents into coherent passages with each passage corresponding to a specific topic. 
%A well-known example is TextTiling \cite{hearst1993subtopic,hearst1993texttiling}. 
%This approach is not popular because identifying topic drift is hard and computationally expensive.
Despite its intuitive motivation, this approach is not widely used because identifying topic drift in documents is hard and computationally expensive. 
%A more appealing method for passage extraction is to extract passages with fixed length windows~\cite{zobel1995efficient}. %\cite{callan1994passage,kaszkiel1997passage,zobel1995efficient}. 
%Window-based passages consist of a fixed number of words, which are independent of text structures. 
Instead, the most widely used methods extract passages with overlapped or non-overlapped windows~\cite{zobel1995efficient}. %, which can be either overlapped or non-overlapped.
%As far as we know, they are the most widely used methods for passage-based retrieval models.
%It is very appealing and popular because of its simplicity.
%Two popular methods in this category are passages extraction with non-overlapped windows and half-overlapped windows. 
%There are also algorithms that dynamically change the start point of window-based passages \cite{kaszkiel2001effective}, but no significant difference has been observed between these window-based approaches in retrieval tasks.                

%paragraph based document ranking
%\subsection{Passage Model}
\textbf{\textit{Passage-based Retrieval Model}}.
%There is a variety of passage models in sentence and answer passage retrieval including feature-based models \cite{yang2016beyond} and neural network based models \cite{severyn2015learning} . However, 
%Due to the lack of passage-level annotations, most passage-based retrieval models use statistical models based on unigrams.
Most passage-based retrieval models in previous studies are unigram models constructed on window-based passages with fixed length.
Liu and Croft \cite{liu2002passage} applied the language modeling approach \cite{ponte1998language} on overlapped-window passages and ranked documents with their maximum passage language score. 
Bendersky and Kurland \cite{bendersky2008utilizing} combined passage-level language models with document-level language models and weighted them according to the homogeneity of each document. 
To the best of our knowledge, our work is the first study that incorporates a neural network framework for passage-based retrieval models. 
%A direct reason behind the utilization of unigram language models is the lack of passage-level annotation, which is usually true in ad-hoc retrieval tasks. 
  

%% file: model.tex
%!TEX root=ECIR18_NeuralPassageModel.tex

%When $\{g_i|i\in \left[0,\lfloor \frac{n_d}{\tau} \rfloor \right]\}$ is generated with the same convolution filter, it is intuitive to define a uniform distribution for $P(g_i|d)$. 
%However, using the average of passage scores contradicts our motivation in Section~\ref{sec:introduction} and produces poor retrieval performance \cite{liu2002passage,bendersky2008utilizing}. 
%Instead, a winner-take-all strategy that only uses the best passage score is more popular in practice \cite{callan1994passage,kaszkiel2001effective,wilkinson1994effective,liu2002passage,bendersky2008utilizing}:
%\begin{equation}
%Score(q, d) = \max_{g_i} Score(q,g_i)
%\end{equation}
%In our proposed neural network framework, this strategy is implemented with a max-pooling layer over the convolution layer on the matching matrix $M(q,d)$.

%probability of $q$ given the unigram language model of $g$ ($M_g$)

\section{Neural Passage Model}
\label{sec:neural_model}

\begin{wrapfigure}[13]{r}{0.57\textwidth}
	%\begin{figure}[t]
	\centering
	\includegraphics[width=3in]{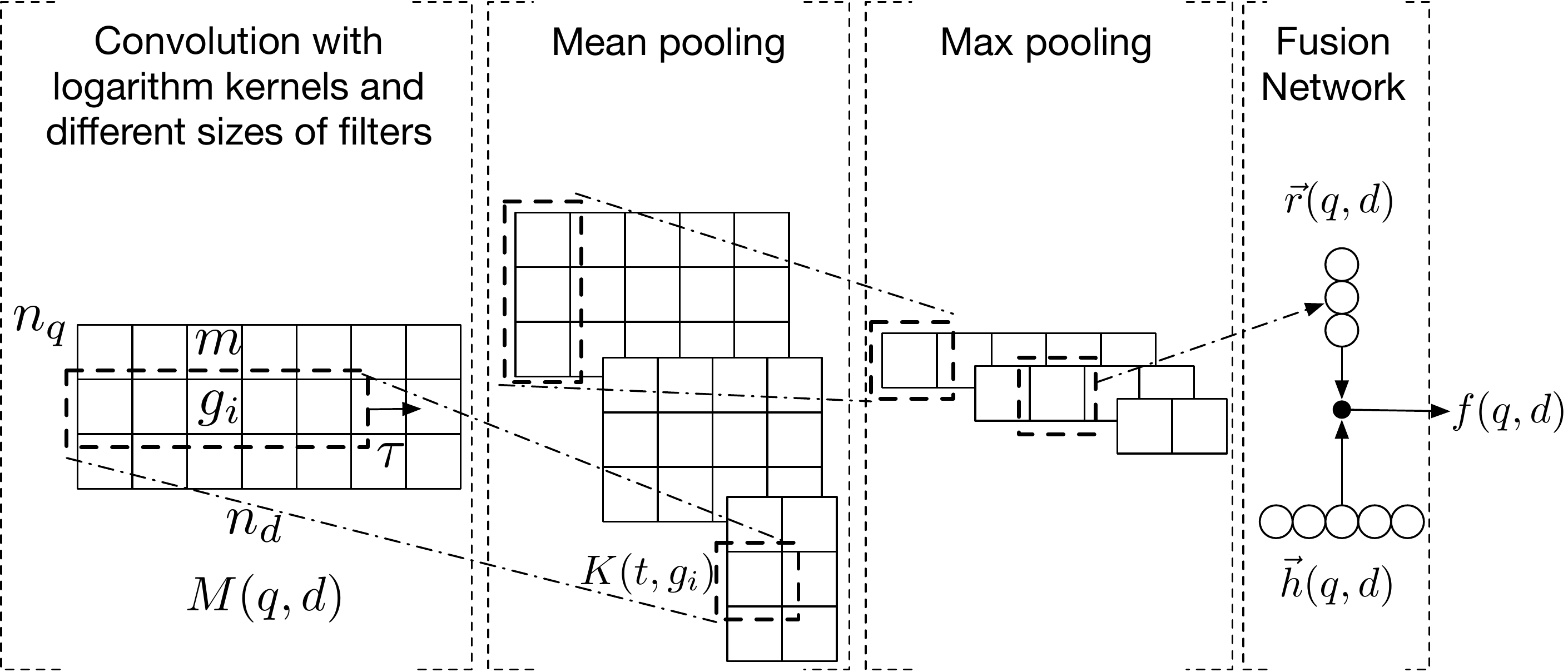}
	\caption{The structure of the NPM. %$n_d$, $n_q$ are the length of document $d$ and query $q$; $M(q,d)$ is a matching matrix of terms in $q$ and $d$; $m$, $\tau$, $K(t,g_i)$ are the window size, the step size and the output of a convolution filter on term $t$ and passage $g_i$ ; $\vec{r}(q,d)$ is a vector of scores from passages extracted with different filters; $\vec{h}(q,d)$ is the query/document features, and $f(q,d)$ is the final ranking score for $q$ and $d$.
	}
	\label{fig:npm}
	%\end{figure}
\end{wrapfigure}

In this section, we describe how to formulate the passage-based document ranking with a neural framework and aggregate information from passages with different granularities in our neural passage model (NPM).
The overall model structure is shown in Figure~\ref{fig:npm}.

\textbf{Passage-based Document Ranking}.
Passage-based retrieval models use passages as representatives for each document and rank documents according to their passage scores.
%As discussed above, most passage-based retrieval models use window-based passages and the state-of-the-art methods compute passage scores by applying the language modeling approach on each passage~\cite{liu2002passage,bendersky2008utilizing}.
Specifically, given a query $q$ and a passage $g$ extracted with a fixed-length window, the score of $g$ is the maximum log likelihood of observing $q$ given $g$'s unigram language model as
\begingroup\makeatletter\def\f@size{8}\check@mathfonts
\begin{equation}
\log P(q|g) = \sum_{t \in q}\log P(t|g)= \sum_{t \in q}\log((1-\lambda_c)\frac{tf_{t,g}}{n} + \lambda_c\frac{cf_t}{|C|})
\label{equ:language_model}
\end{equation}
\endgroup
where $tf_{t,g}$ is the count of $t$ in $g$, $cf_t$ is the corpus frequency of $t$, $|C|$ is the length of the corpus and $\lambda_c$ is a smoothing parameter.

Assuming that passages can serve as proxies of documents~\cite{bendersky2008utilizing}, the ranking score of a document $d$ under the passage-based document ranking framework should be 
%\begin{equation}
$Score(q, d) = \log\sum_{g \in d}P(g|d)\cdot P(q|g)$.
%\end{equation}
Intuitively, $P(g|d)$ could be a uniform distribution since all passages are extracted following the same methodology.
However, averaging passage scores produces poor retrieval performance in practice and the state-of-the-art models adopt a winner-take-all strategy that only uses the best passage to compute document scores~\cite{liu2002passage,bendersky2008utilizing}:  
\begingroup\makeatletter\def\f@size{8}\check@mathfonts
\begin{equation}
Score(q, d) = \max_{g \in d} \log P(q|g)
\label{equ:max}
\end{equation}
\endgroup
%the structure of the neural passage retrieval model. 
%Passage-based retrieval models use passages as representatives for each document and rank documents according to their passage scores.    
%We design a convolution layer on the matching matrix of query-document terms to extract window-based passages and define a logarithm kernel with mean and max pooling to compute passage model scores.
%The information from different sizes of passages is further aggregated with a fusion network that considers both document homogeneity and query characteristics.
%Also, we show that previous passage-based retrieval models based on language modeling approach can be viewed as special cases in our model with predefined parameters.
%The extraction of window-based passages in our model is a convolution process on the matching matrix of query-document terms, and the passage language model used by prior work can be viewed as a logarithm kernel with mean and max pooling in the convolution layer.
%Further, to aggregate information from different sizes of passages, we propose a fusion network that considers both document homogeneity and query characteristics.
%we propose a fusion network that aggregates information from different sizes of passages to form the final ranking scores.  
%The overall structure of the neural passage model is shown in Figure~\ref{fig:npm}.

%\subsection{Convolution and Passage Extraction}
\textbf{\textit{Passage Extraction with a Convolution Layer}}.\label{sec:convolution}
%Window-based passage extraction is widely used for retrieval tasks \cite{zobel1995efficient,liu2002passage,bendersky2008utilizing}. 
%The passage extraction in the neural passage model is achieved with a convolution process.
%Let $n_q$ be the length of query $q$ and $n_d$ be the length of document $d$, then the matching of terms in $q$ and $d$ is a matrix  $M(q,d) \in \mathbb{R}^{n_q\times n_d}$ in which $M(q,d)_{i,j}$ represents the matching between the $i$th term in $q$ and the $j$th term in $d$.
Given a fixed length window, window-based passages are extracted by sliding the window along the document with fixed step size. 
Formally, given a document $d$ with length $n_d$, the set of extracted passages $G(d)$ with window size $m$ and step size $\tau$ is $G(d) = \{ g_i | i \in \left[ 0, \lfloor \frac{n_d}{\tau} \rfloor \right]  \}$
%\begin{equation}
%G(d) = \{ g_i | i \in \left[ 0, \lfloor \frac{n_d}{\tau} \rfloor \right]  \}
%\end{equation}
where $g_i$ represents the $i$th passage starting from the $i\cdot\tau$th term in $d$ with size $m$. 
%When $\tau$ is smaller than $m$, extracted window-based passages have overlaps with their neighbors.
%\begin{figure}
%	\centering
%	\includegraphics[width=3in]{./figure/convolution.pdf}
%	\caption{Passage extraction as a convolution process on the matching matrix of query $q$ and document $d$ ($M(q,d)$) with filter size $m$, stride $\tau$ and kernel function $K$.}
%	\label{fig:convolution}
%\end{figure}
 %(either binary matching or matching with word embeddings). 
%If $q$ has $n_q$ terms and $d$ has $n_d$ terms, then $M(q,d)$ is a $(n_q, n_d)$ dimensional matrix in which the cell $M(q,d)_{i,j}$ represents the matching score between the $i$th term in $q$ and the $j$th term in $d$. 
%Let $M(q,d)$ be the matching matrix of terms in query $q$ and document $d$, and $n_q$ be the length of $q$, then $M(q,d)$ is a $\mathbb{R}^{n_q\times n_d}$ dimensional matrix in which $M(q,d)_{i,j}$ represents the matching of the $i$th term in $q$ and the $j$th term in $d$. 
Let $n_q$ be the length of query $q$, then the matching of terms in $q$ and $d$ is a matrix  $M(q,d) \in \mathbb{R}^{n_q\times n_d}$ in which $M(q,d)_{i,j}$ represents the matching between the $i$th term in $q$ and the $j$th term in $d$.
%In the simplest case, the matching score is a binary variable (1 if the two terms are same and 0 if they are not), but it could also be other value such as the cosine similarity between the term's embedding representations. 
In this work, we define the matching of two terms as a binary variable (1 if the two terms are same and 0 if they are not).
%The matching of two terms can either be scalar (i.e. the cosine similarity between word embeddings) or binary (1 if the two terms are same and 0 if they are not).
%Let $K(q,g_i)$ be the score of passage $g_i$ given $q$, then the sliding window used in window-based passage approach can be  a convolution filter with size $m$, stride $\tau$ and kernel function $K$ over $M(q,d)$, which is shown in the left hand side of Figure~\ref{fig:npm}. 
Let $K(t,g_i)$ be the score of $g_i$ given term $t$ in $q$, then the extraction of window-based passages can be achieved with a convolution filter with size $m$, stride $\tau$ and kernel $K$ over $M(q,d)$. % (the left-hand side of Figure~\ref{fig:npm}).
Passages with different granularities can be directly extracted with different sizes of filters. 
%The extraction of window-based passages is actually a convolution process.

\iffalse
%The bridge between window-based passages and convolution is important because convolution is a standard process in neural networks. 
%From this perspective, the passage extraction of previous passage-based retrieval models~\cite{liu2002passage,bendersky2008utilizing} can be viewed a convolution layer with fixed filters in our neural network framework.
%By formulating a convolution layer, we can derive previous passage-based retrieval models as special cases in our neural network framework, which will be discussed later in this section.
\fi

%\subsection{Language Model as a Logarithm Kernel}
\textbf{\textit{Language Modeling as a Logarithm Kernel}}.\label{sec:language_model}
Let $M(t,g_i)$ be the binary matching matrix for term $t$ and $g_i$ with shape $(1,m)$ and $W \in \mathbb{R}^{m}$, $b_t\in \mathbb{R}$ be the parameters for a logarithm convolution kernel $K$, then we define the passage model score for $q$ and $g_i$ as
\begingroup\makeatletter\def\f@size{8}\check@mathfonts
\begin{equation}
Score(q, g_i) = \sum_{t \in q}K(t,g_i) = \sum_{t \in q}\log(W\cdot M(t,g_i)  + b_t)
\label{equ:kernel}
\end{equation}
\endgroup
%where $W \in \mathbb{R}^{m}$ and $b_t\in \mathbb{R}$ are parameters for the logarithm convolution kernel $K$.

%Now we show how to derive the passage-based document ranking framework with our neural model. 
Let $W$ be a vector of 1 and $b_t$ be $\frac{\lambda_c \cdot m \cdot  cf_t}{(1-\lambda_c) |C|}$, then $K(t,g_i)$ is equal to $\log P(t|g_i)$ in Equation~\ref{equ:language_model} plus a bias ($\log \frac{m}{1-\lambda_c}$). 
Thus, the term-based language modeling approach can be viewed as a logarithm activation function over the linear projection of $M(t,g_i)$.
%where $W$ is a constant matrix with elements equal to 1, and $b_t$ is a bias term equal to $\frac{\lambda_c \cdot m \cdot  cf_t}{(1-\lambda_c) |C|}$.
Further, if we implement the sum of query term scores with a mean pooling and the winner-take-all strategy in Equation~\ref{equ:max} with a max pooling%(the middle part of Figure~\ref{fig:npm})
, then the passage-based document ranking framework can be completely expressed with a three-layer convolution neural network. % with a logarithm kernel and predefined parameters. 

\textbf{\textit{Aggregating Passage Models with a Fusion Network}}.
Bendersky and Kurland observed that the usefulness of passage level information varies on different documents~\cite{bendersky2008utilizing}. 
To consider document characteristics, they proposed to combine the passage models with document models using document homogeneity scores $h^{[M]}(d)$ as
%\begin{equation}
$Score(q, d) = h^{[M]}(d)P(q|d) + (1 - h^{[M]}(d))\max_{g \in d}P(q|g)$
%\label{equ:homo_fusion}
%\end{equation}
where $h^{[M]}(d)$ could be length-based ($h^{\left[ length \right] }$), entropy-based ($h^{\left[ ent \right] }$), inter-passage  ($h^{\left[ intPsg \right] }$) or doc-passage  ($h^{\left[ docPsg \right] }$):
\begingroup\makeatletter\def\f@size{8}\check@mathfonts
\begin{equation}
\begin{split}
h^{\left[ length \right]}(d)  &= 1 - \frac{\log n_d - \min_{d_i \in C}\log n_{d_i}}{\max_{d_i \in C}\log n_{d_i} - \min_{d_i \in C}\log n_{d_i}}\\
h^{\left[ ent \right]}(d)  &= 1 + \frac{\sum_{t'\in d}P(t' |d)\log (P(t' |d))}{\log n_d}\\
h^{\left[ intPsg \right]}  = \frac{2}{\lceil \frac{n_d}{\tau} \rceil(\lceil \frac{n_d}{\tau} \rceil -1)}&\sum_{i < j;g_i, g_j \in d}\cos(g_i,g_j), ~~~~ h^{\left[ docPsg \right]}  = \frac{1}{\lceil \frac{n_d}{\tau} \rceil}\sum_{g_i \in d}\cos(d,g_i)
\end{split}
\label{equ:homogeneity}
\end{equation}
\endgroup
where $\cos(d, g_i)$ is the cosine similarity between the tf.idf vector of $d$ and $g_i$.

Inspired by the design of homogeneity scores and studies on query performance prediction~\cite{zhao2008effective}, we propose a fusion network that aggregates scores from passages according to both document properties and query characteristics. 
We extract features for queries and concatenate them with the homogeneity features to form a fusion feature vector $\vec{h}(q,d)$.
For each query term, we extract their inverse document/corpus frequency and a clarity score~\cite{zhao2008effective} defined as 
%\begin{equation}
%\begin{split}
%IDF_t &= \frac{\log\frac{|D|+0.5}{D_t}}{|D|+1}, ~~~~ICF_t = \log\frac{cf_t}{|C|} \\
$SCQ_t = (1 + \log(cf_t))\log(1 + idf_t)$
%\end{split}
%\label{equ:query_features}
%\end{equation}
where $idf_t$ is the inverse document frequency of $t$. 
For each feature, we compute the sum, standard deviation, maximum, minimum, arithmetic/geometric/harmonic mean and coefficient of variation for $t \in q$.
We also include a list feature as the average scores of top 2,000 documents retrieved by the language modeling approach.
Suppose that $\vec{h}(q,d)\in\mathbb{R}^\beta$ and let $\vec{r}(q,d)\in\mathbb{R}^\alpha$ be a vector where each dimension denotes a score from one convolution filter, then the final ranking score $f(q,d)$ is computed as 
\begingroup\makeatletter\def\f@size{8}\check@mathfonts
\begin{equation}
\begin{split}
Score(q,d) = f(q,d) &= \tanh\big(\vec{r}(q,d)^T\cdot \phi(\vec{h}(q,d))  + b_R\big) 
%\phi(\vec{h}(q,d))_i &= \frac{\exp (\vec{W_R^i}\cdot \vec{h}(q,d))}{\sum_{j=1}^{\alpha}\exp (\vec{W_R^j}\cdot \vec{h}(q,d))}
\end{split}
\label{equ:fusion}
\end{equation} 
\endgroup
where $\phi(\vec{h}(q,d))=\frac{\exp (\vec{W_R^i}\cdot \vec{h}(q,d))}{\sum_{j=1}^{\alpha}\exp (\vec{W_R^j}\cdot \vec{h}(q,d))}$ and $W_R \in \mathbb{R}^{\alpha \times \beta}$, $b_R\in \mathbb{R}$ are parameters learned in the training process.

%% file: methodology.tex
%!TEX root=ECIR18_NeuralPassageModel.tex
\section{Experiment and Results}\label{sec:exp}

In this section, we describe our experiments on Robust04 with 5-fold cross validation~\cite{huston2014comparison}. % to show the effectiveness of the neural passage model.
For efficient evaluation, we conducted an initial retrieval with the query likelihood model \cite{ponte1998language} and performed re-ranking on the top 2,000 documents.
We reported MAP, NDCG@20, Precision@20 and used Fisher randomization test~\cite{smucker2007comparison} ($p<0.05$) to measure the statistical significance. 
Our baselines include the max-scoring language passage model~\cite{liu2002passage}(MSP[base]) and the state-of-the-art passage-based retrieval model with passage weighting~\cite{bendersky2008utilizing} -- the MSP with length scores (MSP[length]), the MSP with entropy scores (MSP[ent]), the MSP with inter-passage scores (MSP[interPsg]) and the MSP with doc-passage scores (MSP[docPsg]).
%\textbf{\textit{Parameter Settings}}.
%The MSP models require a predefined passage size to extract passages from documents. 
We follow the same parameter settings used by Bendersky and Kurland \cite{bendersky2008utilizing} and tested all models with passage size 50 and 150 separately.
%For our neural passage model, we use the convolution filters and kernels described in Section~\ref{sec:neural_model}. 
%One of the advantages of our neural passage model is its ability to simultaneously extract and weight passages with multiple sizes.
%However, to ensure a fair comparison, we only use filters with length 50, 150 and ``infinity".
We used filters with length 50, 150 and $\infty$ for NPMs and set $\tau$ as the half of the filter lengths.
The filter with length 50 extracts the same passages used in MSP models with passage size 50, and the filter with length $\infty$ treats the whole document as a single passage.
Notice that the MSP with passage weighting~\cite{bendersky2008utilizing} also uses sizes 50 (or 150) and $\infty$ to combines the scores of passages and the whole document.
We tested the NPMs with document homogeneity features (NPM[doc]), query features (NPM[query]) and both (NPM[doc+query]).
Due to the limit of Robust04, we only have 249 labeled queries, which are far from enough to train a robust convolution kernel with hundreds of parameters.
Therefore, we fixed the convolution kernels as discussed in Section~\ref{sec:neural_model}.
%We used batch size 64 and learning rate 0.1 for stochastic gradient descent. 
%Also, we randomly selected 20\% of the training data as the validation set to choose the best model in the training process.

\iffalse
As far as we know, this is the state-of-art passage-based ad-hoc retrieval model. 

\textbf{Max-scoring language passage model}: As discussed in Section~\ref{sec:language_model} and \ref{sec:fusion}, the MSP[base] model applies the language modeling approach (as shown in Equation~\ref{equ:language_model}) on passages extracted with fixed length sliding window. The step size is exactly half of the window size and each passage has fifty-percent overlap with passages around it. 
The MSP[base] model rank documents according to their max-score passage.
It can be viewed as a special case of our neural passage model without the fusion layer.

\textbf{Passage language model with document homogeneity}: Bendersky and Kurland extended MSP[base] by incorporating a document homogeneity score to weightly combine passage language models with document language models. 
With different homogeneity score (as described in Section~\ref{sec:fusion}), there are four variations of this model: MSP with length score (MSP[length]), MSP with entropy score (MSP[ent]), MSP with inter-passage score (MSP[interPsg]) and MSP with doc-passage score (MSP[docPsg]).
We tested each variation of MSP individually and reported the results in Section~\ref{sec:result}.
\fi

%% file: experiment_results.tex
%!TEX root=ECIR18_NeuralPassageModel.tex
%\section{Results}
%\label{sec:result}
%In this section, we report the overall performance of different passage-based retrieval models on Robust04 and analyze the parameters of our neural passage model to gain more insights on its ranking process.
\textbf{\textit{Overall Performance}}.
\begin{table}[t]
	\centering
	\caption{The performance of passage-based retrieval models. $*$, $+$ means significant differences over MSP[base] and MSP[docPsg] with passage size (150, $\infty$) respectively. %The best performance is highlighted in boldface.
	}
\scalebox{0.7}{
	\begin{tabular}{ p{2.8cm} || p{2cm} | p{2cm} | p{2cm} || p{2cm} | p{2cm} | p{2cm}   } %p{5mm}
		\hline
		%\multicolumn{1}{c||}{ } & \multicolumn{6}{c}{GOV2 collection}\\ \hline 
		& MAP & NDCG@20 & Precison@20 & MAP & NDCG@20 & Precison@20 \\\hline\hline
		\multicolumn{1}{c||}{} & \multicolumn{3}{c||}{Passage Size (50, $\infty$)} & \multicolumn{3}{c}{Passage Size (150, $\infty$)}\\ \hline 
		$\!\!$MSP[base]  & 0.193 & 0.317 & 0.288 & 0.207 & 0.335 & 0.302 \\ \hline \hline
		$\!\!$MSP[length]  & 0.210 & 0.333 & 0.298 & 0.223$^*$ & 0.355$^*$ & 0.315$^*$ \\ \hline
		$\!\!$MSP[ent]  & 0.209 & 0.338 & 0.304 & 0.216$^*$ & 0.349$^*$ & 0.314$^*$ \\ \hline
		$\!\!$MSP[interPsg]  & 0.204 & 0.329 & 0.296 & 0.215$^*$ & 0.346$^*$ & 0.310$^*$ \\ \hline
		$\!\!$MSP[docPsg]  & 0.206 & 0.331 & 0.296 & 0.226$^*$ & 0.362$^*$ & 0.312$^*$ \\ \hline \hline
		%LDA-LM & 0.292 & 0.405 & 0.504 & 0.244 & \textbf{0.375} & 0.467 \\ \hline
		%PV-DBOW-LM &  \\ \hline
		\multicolumn{1}{c||}{} & \multicolumn{6}{c}{Passage Size (50, 150, $\infty$)} \\ \hline 
		$\!\!$NPM[doc] & 0.255$^{*+}$ & 0.412$^{*+}$ & 0.366$^{*+}$ & - & - & - \\ \hline
		$\!\!$NPM[query] & \textbf{0.256}$^{*+}$ & \textbf{0.416}$^{*+}$ & \textbf{0.369}$^{*+}$ & - & - & - \\ \hline
		$\!\!$NPM[doc+query] & 0.255$^{*+}$ & 0.413$^{*+}$ & 0.367$^{*+}$ & - & - & - \\ \hline
		
	\end{tabular}
}
	\label{tab:results}
\end{table}
Table~\ref{tab:results} shows the results of our baselines and the NPM models with passage size 50 and 150. 
As we can see, the variations of MSP significantly outperformed MSP[base] with the same passage size, and the MSP models with passage size 150 performed better than MSP with passage size 50.
%We followed the experiment framework from Bendersky and Kurland \cite{bendersky2008utilizing} and reported results with passage size 50 and passage size 150.
%Table~\ref{tab:results} also shows the results of NPM with passage size 50 and 150.
Compared to MSP models, the NPM models showed superior performance on all reported metrics. 
%The MAP improvement for the best NPM (NPM[query]) over the best MSP (MSP[docPsg] with passage size 150) is 13.3\%.
As discussed in Section~\ref{sec:neural_model}, MSP models can be viewed as special cases of the NPM with predefined parameters.
With passage size 50, the MSP[base] model is actually a NPM model with filter length 50 and no fusion layer; and the MSP with homogeneity weighting is a NPM model with filter lengths 50, $\infty$ and a linear fusion with document homogeneity scores. 
From this perspective, the NPM model is more powerful than MSP models as it automatically learns to weight passages according to document/query features.
%Our experiments showed that this flexibility is beneficial for passage-based retrieval models in ad-hoc retrieval tasks.

\iffalse
\begin{figure*}[t]
	\centering
	\begin{subfigure}{.3\textwidth}
		\centering
		\includegraphics[width=2in]{./figure/NPM[query]_weight.pdf}
		\caption{NPM[query]}
		\label{fig:npm_query}
	\end{subfigure}%
	\begin{subfigure}{.3\textwidth}
		\centering
		\includegraphics[width=2in]{./figure/NPM[doc]_weight.pdf}
		\caption{NPM[doc]}
		\label{fig:npm_doc}
	\end{subfigure}
	\begin{subfigure}{.3\textwidth}
		\centering
		\includegraphics[width=2in]{./figure/NPM[doc+query]_doc-weight.pdf}
		\caption{NPM[doc+query]}
		\label{fig:npm_doc_query}
	\end{subfigure}
	\caption{The means and standard deviations of passage weights over all query-doc pairs in NPM[query], NPM[doc] and NPM[doc+query].}
	\label{fig:weight}
\end{figure*}
\fi

\begin{wrapfigure}[12]{r}{0.4\textwidth}
	\centering
	\vspace{-5pt}
	\includegraphics[width=2in]{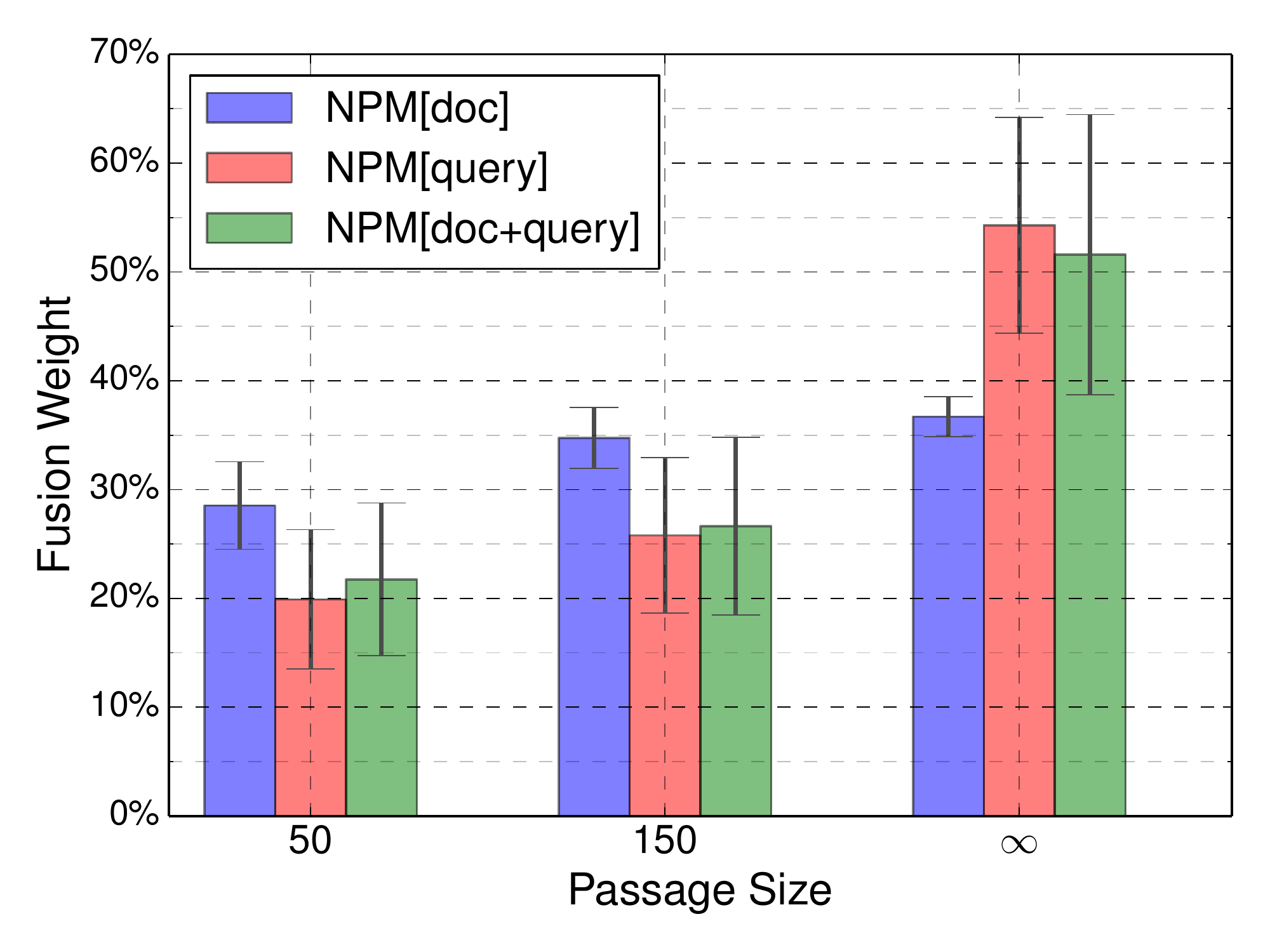}
	\vspace{-30pt}
	\caption{The fusion weights for passages in the NPM averaged over query-doc pairs.}
	\label{fig:weight}
\end{wrapfigure}

\textbf{\textit{Weights of Passages}}.
%We visualized the output weights of our fusion network in Figure~\ref{fig:weight}.
Figure~\ref{fig:weight} shows the means of passage weights $\phi(\vec{h}(q,d))$ on all query-doc pairs for NPM[query], NPM[doc] and NPM[doc+query]. 
%The horizontal axis represents the passage sizes used for different convolution filters.
In our experiments, the passages with size $\infty$ are the most important passages in the NPMs, but the scores from smaller passages also impact the final ranking.
%They weight more than 35\% in the fusion networks of NPM[doc], NPM[query] and NPM[doc+query]. 
%The passages with size 50 and 150 take 20\% to 30\% and 25\% to 35\% of the weights in the NPM models.
%A similar phenomenon can also be observed in NPM[doc+query].
%However, in NPM[doc], the weights for passage size 50 and 150 are approximately the same as the weight of $\infty$.
%In Figure~\ref{fig:weight}, the fusion weights for passages with size 50, 150 and $\infty$ are 28.6\%, 34.7\% and 36.7\%.
%According to our experiments, NPM models with different sets of features can learn different weighting schemes.
Although the MAP of the NPM[query] and NPM[doc] are close, their passage weights are different and they performed differently on 211 of 249 queries on Robust04. 
This indicates that, when evaluating a document with respect to multiple queries, all passages could be useful to determine the document's relevance;
when evaluating multiple documents with respect to one query, models with $\infty$ passage size are more reliable in discriminating relevant documents from irrelevant ones.

%% file: conclusion.tex
%!TEX root=ECIR18_NeuralPassageModel.tex
\section{Conclusion and Future Work}\label{sec:conclusion}

In this paper, we proposed a neural network model for passage-based ad-hoc retrieval. 
We view the extraction of passages as a convolution process and develop a fusion network to aggregate information from passages with different granularities.
Our model is highly expressive as previous passage-based retrieval models can be incorporated into it as special cases. % for it with predefined network parameters.
Due to the limit of our data, we used binary matching matrix and deprived the freedom of the NPM to learn convolution kernels automatically. 
We will explore its potential to discover new matching patterns from more complex signals and heterogeneous datasets. %(e.g. the match between word embeddings) and try more heterogeneous datasets like ClueWeb in future work.
%to produce effective ranking models for ad-hoc retrieval. 
%Due to the limitation of labeled data, we fixed the parameters of convolution kernels in the neural passage model, but we believe that it has potential to automatically discover new matching patterns and produce better retrieval models 

%The proposed neural passage model has good flexibility as it can simultaneously incorporate passages with different granularities and learn to weight them in the learning process.
%Experiments show that our neural passage model indeed outperformed the state-of-art passage-based retrieval models on standard ad-hoc retrieval tasks.
%The effectiveness of our neural passage model indicates that passage information is indeed helpful for the evaluation of topic relevance between queries and documents. \\

%Note that all of the filters in our current neural passage model have width 1, which means that we only consider unigram information. 
%One direction of future work could explore more complicated filters so that we could capture more information such as bigram and trigram. 